\documentclass[aps,pra,preprint,amsmath,amssymb]{revtex4}
\usepackage{graphicx,epsfig,epsf}
\usepackage{dcolumn}
\usepackage{bm}

\begin{document}
\newcommand{\ri}{{\rm i}}
\newcommand{\re}{{\rm e}}
\newcommand{\bx}{{\bf x}}
\newcommand{\bd}{{\bf d}}
\newcommand{\br}{{\bf r}}
\newcommand{\bk}{{\bf k}}
\newcommand{\bE}{{\bf E}}
\newcommand{\bR}{{\bf R}}
\newcommand{\bM}{{\bf M}}
\newcommand{\bn}{{\bf n}}
\newcommand{\bs}{{\bf s}}
\newcommand{\tbs}{\tilde{\bf s}}
\newcommand{\rSi}{{\rm Si}}
\newcommand{\beps}{\mbox{\boldmath{$\epsilon$}}}
\newcommand{\rg}{{\rm g}}
\newcommand{\tr}{{\rm tr}}
\newcommand{\xmax}{x_{\rm max}}
\newcommand{\ra}{{\rm a}}
\newcommand{\rx}{{\rm x}}
\newcommand{\rs}{{\rm s}}
\newcommand{\rP}{{\rm P}}
\newcommand{\up}{\uparrow}
\newcommand{\down}{\downarrow}
\newcommand{\hc}{H_{\rm cond}}
\newcommand{\kb}{k_{\rm B}}
\newcommand{\cI}{{\cal I}}
\newcommand{\tit}{\tilde{t}}
\newcommand{\cE}{{\cal E}}
\newcommand{\cC}{{\cal C}}
\newcommand{\Ubs}{U_{\rm BS}}
\newcommand{\qq}{{\bf ???}}
\sloppy

\title{Distribution of interference in random quantum algorithms} 
\author{Ludovic Arnaud and Daniel Braun}
\affiliation{Laboratoire de Physique Th\'eorique, IRSAMC, UMR 5152 du CNRS,
  Universit\'e Paul Sabatier, 118, route de  
  Narbonne, 31062 Toulouse, FRANCE} 

\begin{abstract}
We study the amount of interference in random quantum algorithms using a
recently derived quantitative measure of interference. To this end we
introduce two random circuit ensembles composed of random sequences of quantum
gates from a universal set, mimicking quantum algorithms in the quantum
circuit representation. We show numerically that these ensembles converge to
the well--known circular unitary ensemble (CUE) for general complex quantum
algorithms, and to the Haar orthogonal ensemble (HOE) for real quantum
algorithms. We provide exact analytical formulas for the average and typical
interference in the circular ensembles, and show that for sufficiently large
numbers of qubits a random 
quantum algorithm uses with probability close to one an amount of
interference approximately equal to the dimension of the Hilbert space. As a
by-product, we offer a new way of efficiently constructing random operators
from the Haar measures of 
CUE or HOE in a high dimensional Hilbert space using universal sets of
quantum gates.    
\end{abstract}
\maketitle

\section{Introduction} It
is generally acknowledged \cite{Bennett00} that quantum information
processing differs from classical information processing fundamentally in
two ways: the use of quantum entanglement and the use of
interference. While
quantum entanglement is undeniably \cite{Curty04} of crucial importance in
tasks like 
quantum teleportation \cite{Bennett93}, and has evolved into a scientific 
field of its own (see \cite{Lewenstein00} for a recent review),
its role in quantum algorithms is less 
clear. Large amounts of entanglement are 
necessarily produced in any quantum algorithm that provides a speed-up over
its classical analogue, but it remains to be seen if the entanglement is a
by-product rather than the fundamental basis of the quantum speed-up
\cite{Jozsa03}. 

Interference on the other hand has received comparatively little attention
in the 
context of quantum information processing. It has been used for a
long time to test the coherence of quantum mechanical propagation
\cite{Ramsey60,Brune96,Vion02}, and it has been proposed as a tool to 
create entanglement between distant atoms \cite{Cabrillo99}, but its role in
complexity theory is virtually unexplored \cite{Beaudry05}.

In contrast to entanglement, which 
is a property of quantum states, interference characterizes the propagation
of states. A quantitative measure of interference was introduced very
recently in \cite{Braun06}. It was shown that a Hadamard gate creates one basic
(logarithmic) unit of interference (an ``i--bit''). Basically all known
useful quantum 
algorithms, including Shor's and Grover's algorithms \cite{Shor94,Grover97}
start off with massive interference by applying Hadamard gates to all
qubits. However, the two algorithms differ substantially in the amount of
interference used in their remaining non--generic part: while the factoring
algorithm uses an exponential amount of interference also for that part
(in fact a number of i--bits close to the number of qubits), only about 3
i--bits suffice for the rest of the 
search algorithm, and that number is asymptotically independent of the
number of qubits. 

The existence of an interference measure makes it
meaningful for the first time to ask the following questions: How much
interference is there typically in a quantum algorithm running on $n$
qubits? How 
is the interference distributed in an ensemble of quantum algorithms? What is
the average interference, what its variance? Are these values different if
the algorithm has a real representation?

\section{Interference distribution in the circular random matrix ensembles}
In order to talk about the statistics of interference, the ensemble needs to
be specified. It is well known that any quantum algorithm (i.e.~any given
unitary transformation in the tensor product Hilbert space
$\mathbb{C}^{2\otimes n}$) can be approximated with arbitrary precision by a
sequence of quantum gates acting on at most two qubits at the time
\cite{DiVincenzo95,Barenco95a,Sleator95}. More precisely, a universal set of
quantum gates is formed by a fixed $U(4)$ transformation, such as the
controlled--NOT gate (CNOT) acting on two arbitrary qubits, in 
conjunction with the set of all $U(2)$ transformations of any single
qubit. Alternatively, any quantum algorithm may be represented by only real
(i.e.~orthogonal) matrices  at the price of doubling the size of the
Hilbert space, with a universal set of quantum gates consisting of the
Hadamard gate and the Toffoli gate \cite{Shi02,Aharonov03}. 
Without any further prior
knowledge of the quantum algorithm, it is natural to chose algorithms from
Dyson's circular unitary ensemble (CUE) for unitary algorithms,
and from 
the so--called Haar orthogonal ensemble (HOE) for algorithms representable
by an orthogonal matrix \cite{Pozniak98}. CUE corresponds to an ensemble of
unitary matrices which is flat with respect to the Haar measure $d\mu_N(U)$ of the
unitary group $U(N)$; and HOE to an
ensemble of orthogonal matrices which is flat with respect to the Haar
measure $d\mu_N(O)$ of the orthogonal group $O(N)$, where $d\mu_N(O)$ is invariant 
under right and left orthogonal transformations ($d\mu(O)=d\mu(V_1OV_2)$ for
any 
two orthogonal matrices $V_1$ and $V_2$). 
We will provide numerical evidence further below
that CUE and HOE represent more realistic quantum circuits indeed very
well, once the number of quantum gates is large enough.

The measure of interference introduced in \cite{Braun06} reduces in the
case of unitary propagation by a $N\times N$ matrix $U$ with matrix elements $U_{ik}$ in
the computational basis to 
\begin{equation} \label{IU}
{\cal I}(U)=N-\sum_{i,k=1}^N|U_{ik}|^4\,,
\end{equation}
with $0\le {\cal I}(U)\le N-1$. Of the two characteristics of interference,
coherence and superposition of a large number of basis states
(equipartition), only the latter distinguishes different entirely coherent
quantum algorithms 
representable by a unitary matrix. The maximum amount of interference is
reached for any quantum algorithm which spreads out each computational basis
state 
equally over all computational basis states, whereas the interference is
zero for a mere permutation of the computational basis states. 

\begin{figure}
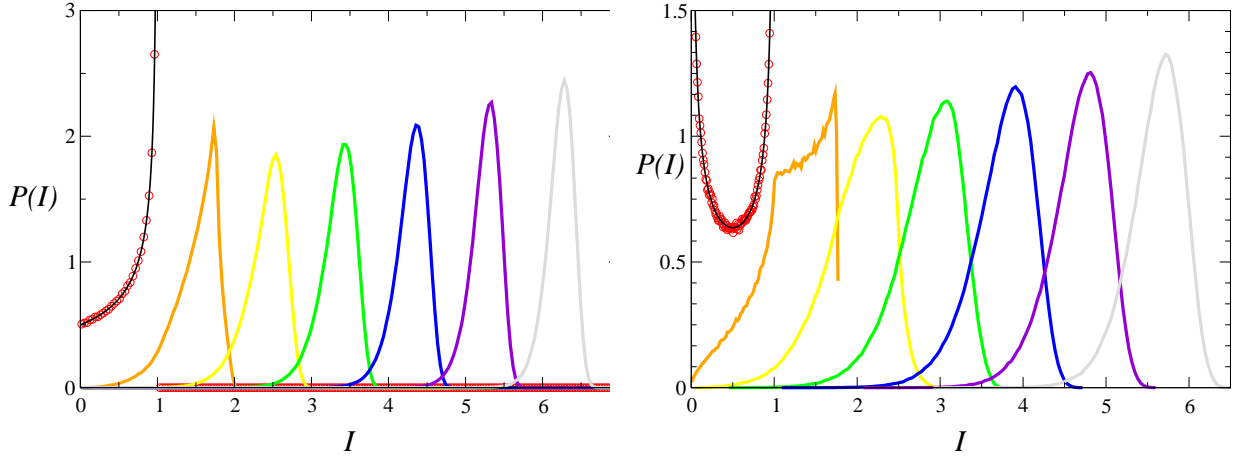

\epsfig{file=CUEinterfstat.eps,width=6cm,angle=270}
\epsfig{file=HOEinterfstat.eps,width=6cm,angle=270}
\caption{(Color online) Distribution of interference in the CUE ensemble
  (left) and HOE ensemble (right) for
  $N=2$ to $N=8$ in steps of 1 (curves from left to right). The full lines for $N=2$ represent
  the analytical results,  
  eqs.(\ref{PU2}) and (\ref{PIO}).  
  }\label{fig.PofIU}        
\end{figure}

We have numerically calculated the distribution of interference $P_{CUE,N}(\cI)$
of $N\times N$ matrices from CUE
using the Hurvitz parametrization for creating large ensembles of random
unitary matrices \cite{Hurwitz1897,Pozniak98}. Figure \ref{fig.PofIU} shows
the result for $N$ between 2 and 8. With growing $N$, the
distribution becomes increasingly peaked on a value close to $N$. For $N=2$
the distribution can be easily calculated analytically. We parametrize $U_2$
with four angles $\alpha$, $\psi$, $\chi$ chosen randomly and uniformly
from the $[0,2\pi [$ and $\varphi=\arcsin(\xi^{1/2})$ with $\xi$ random and
uniform from $[0,1]$,
\begin{equation} \label{U2}
U_2=\re^{\ri \alpha}\left(
\begin{array}{cc}
  \cos\varphi\re^{\ri\psi}&\sin\varphi\re^{\ri\chi}\\
  -\sin\varphi\re^{-\ri\chi}&\cos\varphi\re^{-\ri\psi}\\
\end{array}
\right)\,.
\end{equation}
Thus, $\cI(U_2)=4(\xi-\xi^2)$, and 
\begin{equation} \label{PU2}
P_{CUE,2}(\cI)=\frac{1}{2\sqrt{1-\cI}}\,,
\end{equation}
in very good agreement with the numerical result. 
Fig.\ref{fig.PofIU} indicates that for sufficiently large $N$ all quantum
algorithms will typically contain the same amount of interference of order
$N$. This is confirmed by an exact analytical calculation of the two lowest
moments of the interference distribution. Invariant
integration over the unitary group \cite{Aubert03} gives closed formulas for
integrals of the type
\begin{eqnarray} \label{Z}
Z_{U,N}(m_1,m_2,m_3)&\equiv&
\int
d\mu_N(U)|U_{i_1j_1}|^{2m_1}|U_{i_1j_2}|^{2m_2}|U_{i_2j_2}|^{2m_3}\nonumber\\
&=&\frac{m_1!m_2!m_3!(N-2)!(N-1)!}{(N+m_1-2)!(N+m_3-2)!}\nonumber\\
&&\times\frac{(N+m_1+m_3-2)!}{(N
  +m_1+m_2+m_3-1)!}\,, \nonumber 
\end{eqnarray}
where $d\mu_N(U)$ is normalized to $\int 
d\mu_N(U)=1$, and $i_1,i_2,j_1,j_2$ are arbitrary indices. This leads to the
 average interference
\begin{eqnarray} \label{IavU}
\langle
\cI\rangle_{U,N}&=&N-N^2Z_{U,N}(2,0,0)=
N(1-\frac{2}{N+1})\\
&&\stackrel{N\to\infty}{\longrightarrow}N-2\,,   
\end{eqnarray}
with $\langle\ldots\rangle_{U,N}=\int d\mu_N(U)(\ldots)$.
The 2nd moment can be found from 
\begin{eqnarray} \label{U8av}
\langle
\left(\sum_{i,k}|U_{i,k}|^4\right)^2\rangle_{U,N}
&=&(N(N-1))^2Z_{U,N}(2,0,2)\nonumber\\
&+&2N^2(N-1)Z(2,2,0)\nonumber\\
&+&N^2Z_{U,N}(4,0,0)\nonumber\\
&=& 4\frac{N^2+2N-1}{(N+1)(N+3)} \,.
\end{eqnarray}
Thus, the standard deviation of the interference distribution in the CUE
ensemble,
\begin{equation} \label{sUN}
\sigma_{U,N}=\frac{2}{N+1}\sqrt{\frac{N-1}{N+3}}\,,
\end{equation}
vanishes like $\sim 2/N$ for large $N$. 

Figure \ref{fig.PofIU} also shows the interference distribution
$P_{HOE,N}(\cI)$ for the HOE
ensemble, relevant for quantum algorithms representable with purely real (orthogonal)
$N\times N$ matrices. We constructed this ensemble numerically by
diagonalizing real symmetric matrices drawn from the Gaussian orthogonal
ensemble (GOE) \cite{Mehta91}, which for the relatively small matrix sizes
turned out to be more efficient than Hurvitz's method
\cite{Hurwitz1897,Pozniak98}. Remarkable is the symmetric structure of the 
interference distribution for $N=2$, whose analytical form is easily
obtained from $2\times 2$ rotation
matrices with uniformly distributed rotation angles, 
\begin{equation} \label{PIO}
P_{HOE,2}(\cI)=\frac{1}{\pi\sqrt{\cI(1-\cI)}}\,.
\end{equation}
For $N>4$, the distribution becomes mono-nodal, and more and more peaked with
increasing $N$. 
The method of invariant integration can be generalized to the HOE ensemble
\cite{Braun06c}. The result corresponding to eq.(\ref{Z}) reads 
\begin{eqnarray} \label{Fm}
Z_{O,N}(m_1,m_2,m_3)&\equiv&\int
d\mu_N(O)(O_{i_1j_1})^{m_1}(O_{i_1j_2})^{m_2}(O_{i_2j_2})^{m_3}\nonumber\\
&=&\frac{2^{2-N}\Gamma(\frac{1+m_1}{2})\Gamma(\frac{1+m_2}{2})\Gamma(\frac{1+m_3}{2})}{\pi\Gamma(\frac{N+m_1-1}{2})\Gamma(\frac{N+m_3-1}{2})}\nonumber\\ 
&&\times
\frac{\Gamma(N-1)\Gamma(\frac{N+m_1+m_3-1}{2})}{\Gamma(\frac{N+m_1+m_2+m_3}{2})}\nonumber
\end{eqnarray}
where $m_1,m_2,m_3$ are all even, $\Gamma$ means Euler's gamma function,
and $d\mu_N(O)$ is normalized to $\int 
d\mu_N(O)=1$.
The average interference in the HOE ensemble is then given by
\begin{eqnarray} \label{IavU2}
\langle
\cI\rangle_{O,N}&=&N-N^2Z_{O,N}(4,0,0)=
N(1-\frac{3}{N+2})\nonumber\\
&&\stackrel{N\to\infty}{\longrightarrow}N-3\,,   
\end{eqnarray}
with $\langle\ldots\rangle_{O,N}=\int d\mu_N(O)(\ldots)$. Thus, a real
quantum algorithm of the same size $N$, drawn from HOE, contains on the average
asymptotically slightly less interference than a unitary one drawn from
CUE. However, since the size of the Hilbert space has to be doubled to
express an arbitrary complex algorithm as a real one \cite{Aharonov03},
about twice as much 
interference is needed to run the real algorithm.
The second moment
\begin{eqnarray}
\langle \left(\sum_{ik} (O_{ik})^4\right)^2\rangle_{O,N}&=&
(N(N-1))^2Z_{O,N}(4,0,4) \nonumber\\
&&+2N^2(N-1)Z_{O,N}(4,4,0)\nonumber\\
&&+N^2Z_{O,N}(8,0,0)\nonumber\\
&=&\frac{3N(-4+3N(N+5))}{(N+1)(N+2)(N+6)}\,,
\end{eqnarray}
leads to the variance 
\begin{equation} \label{s2O}
\sigma^2_{O,N}(\cI)=\frac{24N(N-1)}{(N+2)^2(N^2+7N+6)}
\end{equation}
of the interference for the HOE ensemble. Thus, the standard deviation
$\sigma_{O,N}(\cI)$ 
decays as $\sim 2\sqrt{6}/N$ for large $N$, and therefore practically all
algorithms 
drawn from HOE contain for
large $N$ an amount of interference $\cI\sim N-3$. 

\section{Random circuit ensembles}
Shor's algorithm was recently shown to lead
to CUE level statistics \cite{Maity06}, whereas the quantum Fourier transform
alone has a regular spectrum (it is a fourth root of the identity matrix), and
so does Grover's algorithm, which is to 
good approximation a sixth root of the identity matrix \cite{Braun02a}. 
It is therefore natural to ask to what extent are the CUE and the HOE
ensembles representative of realistic quantum algorithms? 

To answer the above question, 
we introduce two random quantum circuit ensembles, the random
unitary circuit ensemble (UCE), and the random orthogonal circuit ensemble
(OCE), constructed to resemble realistic quantum algorithms with $n_g$
randomly chosen gates as follows:
\begin{itemize}
  \item For quantum gate number $i$ ($1\le i\le n_g$), decide whether to
  apply a  one--qubit gate  (with
  probability $p$) or a multi--qubit gate (with probability $1-p$). 
\item If gate $i$ is a  one--qubit gate, chose randomly, uniformly over all
  qubits, and independently from all other gates the qubit on which the
  gate 
  is to act,  and pick as gate a random unitary
  $2\times 2$ matrix from CUE for the
  construction of an UCE algorithm, or the Hadamard gate for building an OCE
  algorithm. 
\item If gate $i$ is a  multi--qubit gate, chose randomly, uniformly over all
  qubits, and independently from all other gates a control qubit (two
  control qubits) and a
  target qubit, and apply the CNOT gate (the Toffoli gate) to these
  qubits for UCE (OCE), respectively.
\item Repeat this procedure for all gates $i=1,\ldots,n_g$ and concatenate the
  obtained gates to form the entire quantum algorithm. 
\end{itemize}
A similar ensemble of random quantum circuits was introduced in
\cite{Emerson03}, where, however, one random constant depth gate was
iterated, and the entangling gate was constructed from simultaneous
nearest neighbor interactions. Nevertheless, according to \cite{Emerson05},
one might expect at least an exponential convergence to CUE (HOE) also for UCE (OCE),
respectively, and this is what we are going to show numerically. 

\begin{figure}
\epsfig{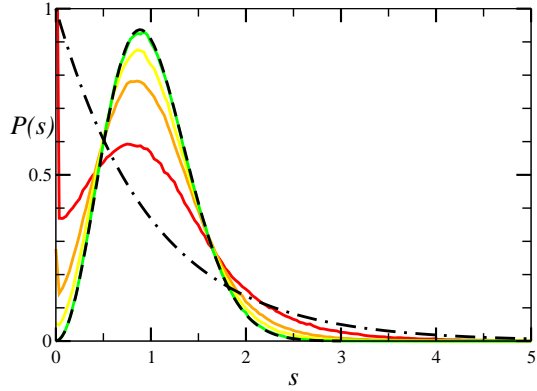}
\caption{(Color online) $P(s)$ for $n_g=10$ (red), 15 (orange) 20 (yellow)
  and 40 
  (green --- increasing values
  of $P(1)$ in this order) for UCE with $n=4$
  and $p=0.5$, $n_r=10^5$ matrices, compared
  to the   Wigner   surmise 
  $P_W(s)$ (dashed black line) and the Poisson result,
  $P(s)=\exp(-s)$ (dashed-dotted black line).\\ }\label{fig.PsU}         
\end{figure}

\begin{figure}
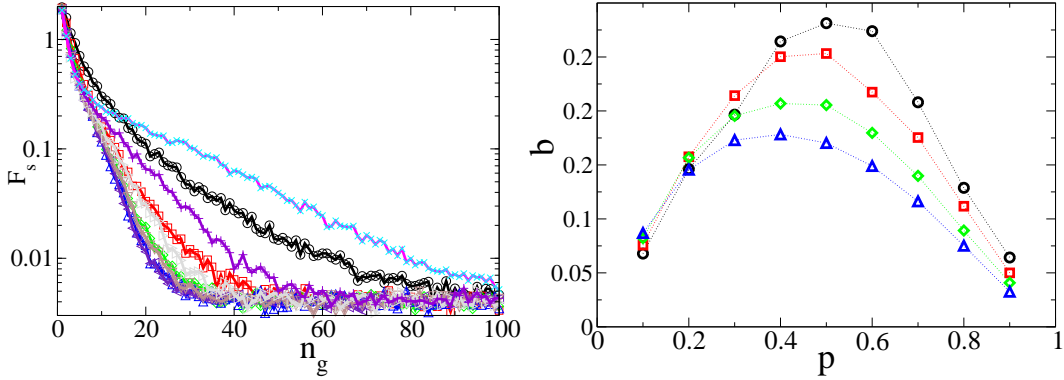

\epsfig{file=FsofnpUCE.eps,width=5cm,angle=270}
\epsfig{file=bofpH_UCEPs.eps,width=5cm,angle=270}
\caption{(Color online) Convergence of $P(s)$ for UCE to 
  $P_W(s)$ as function of 
  $n_g$ for $n=4$ qubits, $n_r=10^3$, and different values of $p$: $p=0.1$ black squares,
  $p=0.2$ red squares, $p=0.3$ green diamonds, $p=0.4$ blue triangles up,
  $p=0.5$ indigo triangles left, $p=0.6$ brown triangles down, $p=0.7$ grey
  triangles right, $p=0.8$ violet pluses, and $p=0.9$ magenta Xs (left). Rate of convergence 
  $b$ as function of the probability $p$ for $n=3$ (circles), $n=4$
  (squares), $n=5$ (diamonds), and $n=6$ (triangles up). \\  }\label{fig.FU}        
\end{figure}

We first examine the convergence by comparing
the distribution $P(s)$ of nearest neighbor spacings $s$ of the eigenphases 
$\varphi_l$ of
the $N\times N$ unitary matrices. For large $N$, and 
average $s$ normalized to unity, 
CUE leads to  a $P(s)$  well approximated by the Wigner
surmise \cite{Mehta91}
\begin{equation} \label{PW}
P_W(s)=\frac{32s^2}{\pi^2}\re^{-4s^2/\pi}\,.
\end{equation}
Deviations are of order $10^{-3}$ \cite{Haake91}. For $n=4$, the minimum
number of gates that leads to an approximately 
constant density of eigenphases, such  
that unfolding the spectrum \cite{Mehta91} is unnecessary, is $n_g\simeq
10$. For even smaller 
numbers of gates strong peaks at $\varphi=0$ and $\varphi=\pi$ arise in the
density of states
corresponding to a predominance of real eigenvalues, but otherwise the
density is already flat. 
Fig. \ref{fig.PsU} shows $P(s)$ for UCE for $n=4$ and several values of
$n_g$ ($n_r=10^5$ realizations). For 
 small $n_g$, $P(s)$ has a strong peak at
 $s=0$. The rest of the distribution is 
 between the Poisson result of uncorrelated phases, $P(s)=\exp(-s)$, and
 the Wigner surmise $P_W(s)$. The peak at $s=0$ becomes smaller
 and smaller as the number of gates increases, and at the same time a  more
 and more pronounced maximum at $s=1$ arises, resulting in a distribution
 which rapidly  approaches the Wigner surmise, eq. (\ref{PW}). For $n_g=40$,
 $P(s)$ is virtually indistinguishable from 
 $P_W(s)$.   
We examine the convergence quantitatively with the help of the quantity
\begin{equation} \label{F}
F_s=\int_0^\infty
\left(\sqrt{P_{UCE}(s)}-\sqrt{P_W(s)}\right)^2\,ds=2\left(1-\int_0^\infty\sqrt{P(s)P_W(s)}\,ds\right)  
\end{equation}
which measures a squared distance between the (square roots of) the level
spacing distributions $P_{UCE}(s)$  of UCE and $P_W(s)$. 
Fig.~\ref{fig.FU} shows $F_s$ as function of $n_g$ for UCE for various
values of $p$ and 
$n=4$ qubits, obtained from $10^3$ random algorithms. For
$p$ different from 0 and 1, $F_s$ decays to a good approximation
exponentially as $\sim \exp(-b(n,p)n_g)$, 
with a rate $b$ that depends on $p$ and $n$, before saturating at a small level largely
independent of $p$. The latter is due to the numerical fluctuations in
$P(s)$ present for any finite $N$, as is easily checked by varying
$N$. The finite precision of $P_W(s)$ for $N>2$ sets another lower bound on
the values of $F$ that can be possible achieved. Fig.~\ref{fig.FU} also
shows that $b(n,p)$, as obtained from a fit
of $\ln F_s$ to a linear function of $n_g$ between $F_2=2$ and $F_2=0.1$ 
has a maximum around $p=0.5$. The convergence rates decrease with
increasing $n$, and the maximum of the convergence rate as function of $p$
shifts to somewhat smaller values of $p$. 

\begin{figure}
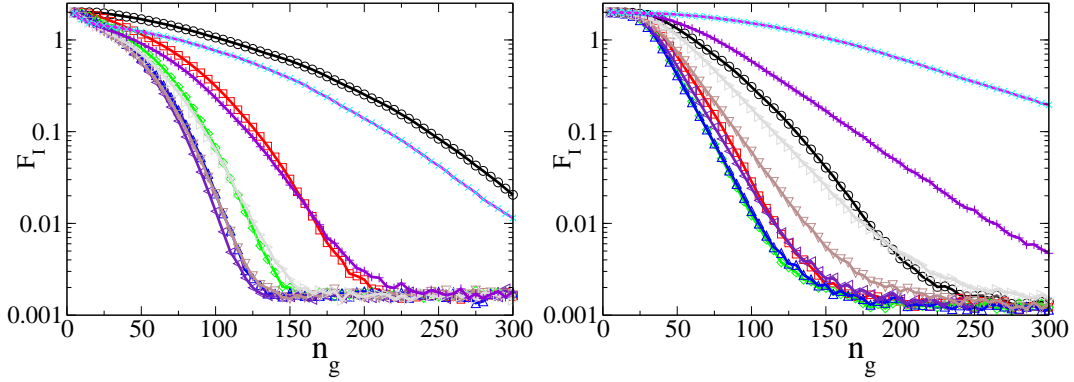

\epsfig{file=FofnpUCE_32.eps,width=5cm,angle=270}
\epsfig{file=FofnpOCE_32.eps,width=5cm,angle=270}
\caption{(Color online) Convergence of $P(\cI)$ for UCE (left) and OCE
  (right) to 
  the interference distribution of CUE and HOE, respectively, as function of 
  $n_g$ for $n=5$ qubits ($N=32$) and different values of $p$. Same symbols as in Fig.\ref{fig.FU}.\\
  }\label{fig.FUI}        
\end{figure}

Numerical evidence presented
in \cite{Pozniak98} indicates the same form of $P(s)$ for HOE as for CUE,
eq.(\ref{PW}), in
particular a quadratic level repulsion $P(s)\propto s^2$ for $s\ll
1$. We have examined the convergence of OCE to HOE based on $P(s)\to P_W(s)$
as well, 
and have found similar results as in Fig.\ref{fig.FU}. However, it is
clearly not 
possible to determine the limiting ensemble based on $P(s)$ alone. We
therefore also examined directly the interference distributions $P(\cI)$ for
both 
random circuit ensembles, as $P(\cI)$ is, after all, what we are interested
in.  
\begin{figure}
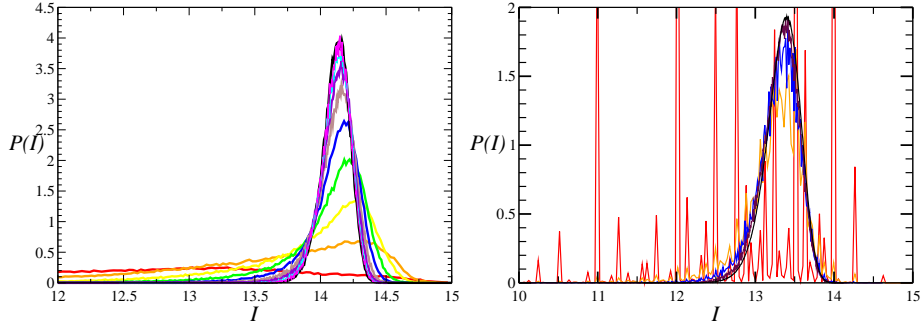

\epsfig{file=UCEinterfstat.eps,width=6cm,angle=0}
\epsfig{file=PofIOCE.eps,width=6cm,angle=0}
\caption{(Color online) Interference distribution $P_{UCE,N}(\cI)$ (left),
  for 
  $n_g=10,20,30,40,50,60,70,80,90,100$, maxima increasing in this order, and 
  $P_{OCE,N}$ (right) for  $n_g=20,50,70,100$     (red, orange, blue, maroon,
  respectively) compared to the $P(\cI)$ of the
  circular random matrix ensembles  (black lines). All curves
   have $n=4$ ($N=16$); $n_r=10^5$ random realizations were used for UCE
  and OCE,  $n_r=10^7$ for CUE  and HOE.   \\
  }\label{fig.PIUCE}        
\end{figure}
Fig.\ref{fig.PIUCE} shows how the interference distribution of UCE
for $n=4$ evolves between $n_g=10$ to $n_g=100$ from a broad flat
distribution to the strongly peaked interference distribution of CUE. The
interference distribution $P_{OCE,N}(\cI)$ for OCE fluctuates much more for a given number of gates
compared to the one for UCE, $P_{UCE,N}(\cI)$, but rapidly approaches $P_{HOE,N}(I)$ as well. To
examine the convergence quantitatively, we
define the quantity $F_I$ as in (\ref{F}), but with $P(s)$ and $P_W(s)$
replaced by the interference distributions 
$P_{UCE,N}(\cI)$ and $P_{CUE,N}(\cI)$ for UCE and CUE (by $P_{OCE,N}(\cI)$ and
$P_{HOE,N}(\cI)$ for OCE and HOE). Note that $P_{\rm CUE}(\cI)$
and $P_{\rm HOE}(\cI)$ have now to be computed numerically as 
well. We did so for the same dimension of Hilbert space $N=2^n$ considered
for UCE and OCE. We used $n_r=10^7$ realizations for
$n\in\{4,5\}$, for both CUE and HOE, as well as for $n=6$, HOE;
$n_r=5\cdot 10^6$ for $n=7$, and $n_r=10^6$ for $n=8$ (HOE); and $n_r=4\cdot
10^5$ for $n=6$ (CUE). The number of   
realizations chosen for the OCE and UCE ensembles was $10^5$, with the
exception of $n_r=4\cdot 10^5$ for $n=6$, UCE. 
 Fig.\ref{fig.FUI} shows the results for  $F_I(n_g)$ for $n=5$. The curves
 for the other values of $n$ examined ($n\in\{4,6,7,8\}$) look very similar,
 but 
  the convergence slows down with increasing $n$. 

\begin{figure}
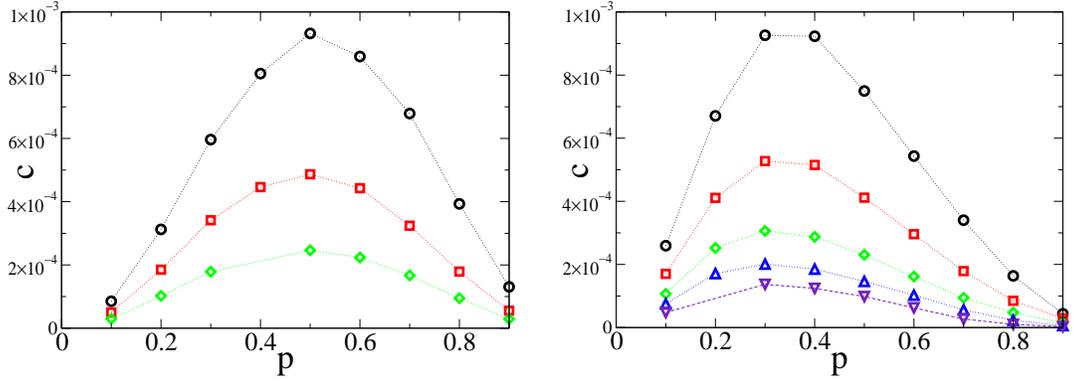

\epsfig{file=cofpH_UCE.eps,width=5cm,angle=270}\hspace{0.5cm}
\epsfig{file=cofpOCE.eps,width=5cm,angle=270}
\caption{(Color online) Dependence of the 
  fitted Gaussian convergence rate $c$ on the probability $p$ for a single qubit gate
   (a random $U(2)$ for UCE (left) and Hadamard for OCE (right)) for various numbers of
   qubits (circles, squares, diamonds for
   $n=4,5,6$), respectively, and in addition triangles up and triangles down
  for $n=7,8$ for UCE.  }\label{fig.FUIexp}        
\end{figure}
$F_I(n_g)$ for both UCE
  and OCE is very well fitted by a Gaussian, at least up to the point where
  the crossover to the saturated behavior occurs. This is in 
contrast to the exponential convergence of $P(s)$. A fit of $\ln F_I(n_g)$
to $a-c(n,p)\,
n_g^2$ in the range $2\ge F_I(n_g)\ge 0.01$ yields
$c\sim 10^{-4}$ for $4\le n\le 6$, with a maximum of $c$ around
$p\sim 0.5$ for both OCE and UCE (see Fig.\ref{fig.FUIexp}). 

The exponential (or even Gaussian) convergence of the random circuit
ensembles to the corresponding circular ensembles as function of the number
of quantum gates provides a new way of 
economically creating random unitary operators with a flat distribution with
respect to the appropriate Haar measure in an 
exponentially large Hilbert space. The method will work on any
quantum computer on which the relevant universal set
of gates is available. The OCE is particularly interesting, as 
the only randomness resides in the indices of the qubits selected as entry
of the 
Hadamard or Toffoli gates. About $2n_g$ such random numbers $i$ in the range
$0\le i\le n-1$ are needed for $p=1/2$. This offers the possibility to
construct truly random operators, with only a logarithmic overhead of
qubits: A small register of 
$n_a\sim 3\log_2n+1$
auxiliary qubits can be brought repeatedly into superposition of all
computational states by applying a Hadamard gate to each qubit; then the
register is measured in the computational basis and gives a random
number. The outcome of one particular qubit (the highest significant, say),
can be used to choose between the Hadamard and Toffoli gates, and the
remaining bits 
specify the qubit(s) on which to act. Obviously, one might as well use the
actual work qubits to generate these random numbers initially and store them
for later use in the quantum algorithm.  

The question remains open whether the method presented here is efficient in
the sense that the number of quantum gates $n_g$ needed for a given
fidelity $F_I$ or $F_s$ increases at most polynomially with $n$. This
would require that 
the exponents $c(n,p)$ and $b(n,p)$ decay no faster than an inverse power of
$n$ for a given $p$. The same
problem was encountered in \cite{Emerson03,Emerson05} and so
far no 
definite answer has been found. In order to address the question
numerically, much larger values of $n$ have to be considered. The
study of the interference distributions is clearly not well suited for this
purpose, as each algorithm (i.e.~a very large dimensional matrix) 
gives only one number.

\section{Summary}
As a summary, we have introduced two ensembles of random quantum 
algorithms, UCE for general unitary algorithms, and OCE for real orthogonal
algorithms. We have provided numerical evidence that these ensembles converge
for sufficiently large numbers of gates and for a finite probability for both
one--qubit and two--qubit gates (or three-qubit gates), to the well--known
random matrix ensembles 
CUE and HOE, respectively, at least in the sense of coinciding level
spacing distributions and interference distributions. One might consider these
ensembles therefore 
as a new way of efficiently creating random unitaries from the
corresponding Haar measure \cite{Emerson03,Emerson05}. The method is
universal in the sense that it runs on any quantum computer with a universal
set of quantum gates. We have calculated
numerical distributions of 
interference over the CUE and HOE ensembles, and have provided exact
analytical formulas for the lowest moments. For large Hilbert
space dimensions $N$, the interference distributions over CUE and HOE are
peaked on their 
average values $\langle \cI\rangle_{U,N} \simeq N-2$ and $\langle \cI\rangle_{O.N}
\simeq N-3$, respectively, with a width that decays
$\propto 1/N$ in 
both cases. Thus, randomly picked unitary quantum algorithms contain with high
probability basically the same exponentially large amount of 
interference $\cI\sim N$. This result is reminiscent of similar findings for
the amount of entanglement in a random quantum state
\cite{Cappellini06}. Grover's search algorithm is therefore remarkably
exceptional in the sense that its non--generic part (i.e.~the part {\em
  after} bringing the computer into a superposition of all computational
states) uses only a small amount of
interference (the whole algorithm including the initial Hadamard gates
produces exponential interference \cite{Braun06}).

{\em Acknowledgments:}
We would like to thank Bertrand Georgeot for interesting discussions, and
CALMIP (Toulouse) for the use of their computers. This
work was supported by the Agence  
National de la Recherche (ANR), project INFOSYSQQ, and the EC IST-FET
project EDIQIP. 
\bibliography{../mybibs_bt}

\end{document}